\documentstyle[12pt]{article}

\begin{document}

\title{\bf
Quantum supergroup structure of 1+1-dimensional
quantum superplane,
its dual and its differential calculus
}

\author{
{\bf
M. EL Falaki\thanks{E-mail:mantrach@fsr.ac.ma}
}\\
\normalsize {Laboratoire de Physique Th\'eorique,  Facult\'e des Sciences,}\\
\normalsize{Universit\'e Mohamed V,
Av. Ibn Battota BP. 1014, Rabat,
Morocco}\\\normalsize{Abdus Salam International Centre for
Theoretical Physics, Trieste, Italy}
 \and
{\bf E. H. Tahri}\thanks{ E-mail:
tahrie@sciences.univ-oujda.ac.ma}\\
\normalsize{Laboratoire de Physique Th\'eorique,  Facult\'e des Sciences,}\\
\normalsize{Universit\'e Mohamed V, Av. Ibn Battota BP. 1014, Rabat, Morocco}\\
\normalsize{Laboratoire de Physique Th\'eorique et de Particules,
Facult\'e des Sciences},\\
\normalsize{Universit\'e Mohammed I,
BP. 524, 60000 Oujda, Morocco}\thanks{Permanent address}.
}
\date{October 1999}
\maketitle
\newcommand{\nco}{\newcommand}
\newfont{\msbm}{msbm10 at 12pt}
\nco{\CC}{\mbox{\msbm C}}
\nco{\ZZ}{\mbox{\msbm Z}}
\nco{\beq}{\begin{equation}}
\nco{\eeq}{\end{equation}
}
\nco{\bra}{\begin{array}}
\nco{\era}{\end{array}}\nco{\p}{\hat}
\nco{\ot}{\otimes}\nco{\nt}{\noindent}
\nco{\bl}{\biggl}\nco{\br}{\biggr}
\nco{\Br}{\Biggr}\nco{\Bl}{\Biggl}
\nco{\cU}{\cal U}\nco{\cA}{\cal A}
\nco{\bgl}{\biggl}\nco{\bgr}{\biggr}\nco{\lag}{\langle} \nco{\rad}{\rangle}
\nco{\lpt}{\longmapsto}
\nco{\be}{\beta}\nco{\ga}{\gamma}\nco{\te}{\theta}
\nco{\de}{\delta}\nco{\De}{\Delta}\nco{\pa}{\partial}
\nco{\ep}{\epsilon}\nco{\vep}{\varepsilon}\nco{\Om}{\Omega}
\nco{\om}{\omega}\nco{\na}{\nabla}
\nco{\Te}{\Theta}
\nco{\prop}{\end{Proposition}}
\nco{\pro}{\begin{Proposition}}
\nco{\leme}{\end{Lemma}}
\nco{\lem}{\begin{Lemma}}
\nco{\eth}{\end{Theorem}}
\nco{\bth}{\begin{Theorem}}

\newtheorem {Definition}{Definition}
\newtheorem {Proposition}{Proposition:}
\newtheorem {theorem}{Theorem}
\newtheorem {Lemma}{Lemma}
\nco{\bca}{\era\eeq}

\begin{abstract}
\baselineskip 0.2truecm{
We show that the $1+1$-dimensional quantum superplane
introduced by Manin is a quantum supergroup according
to the Faddeev-Reshetikhin-Takhtajan approach. We give its supermatrix
element, its corresponding R-matrix and its Hopf structure. This new point of view
allows us, first, to realize its dual Hopf superalgebra starting from
postulated initial pairings. Second, we construct a right-invariant
differential calculus on it
and then deduce the corresponding quantum Lie superalgebra which
as a commutation super-algebra appears
classical, and as Hopf structure is a non-cocommutative $q$-deformed one.
An isomorphism between the latter and the dual one obtained in
the first method is given.}
\end{abstract}

\pagebreak
\section{Introduction.}
In recent years there has been much interest in the concept
of quantum groups and quantum algebras \cite{Drin} \cite{Jimb}
\cite{FRT} \cite{Wor87C}. Originating from investigations on the quantum
inverse scattering methods and the Yang-Baxter equations, the quantum
groups have found various applications in theoretical physics such as
integrable fields theories, statistical models and conformal field theories
in two-dimensions (see for example \cite{Dobn} \cite{CFZ} \cite{LPS}).

Another major field starting in the seventies and still attracting a lot of
attention among physicists and mathematicians is the field of super-symmetry.
The extension of the activity on quantum groups to this field was started
with the paper of Manin \cite{Man89}, where the multi-parametric quantum
deformation of the supergroup $GL(m/n)$ was introduced. Since then intensive
investigations concerning both structures and representations of the quantum
supergroups were completed (see for example \cite{dota} \cite{Nky98} and
references therein).

In Ref \cite{Man89}, Manin extended the notion of quantum space \cite{Man88}
to that of quantum super-space, called also quantum super-plane, of which
the defining quadratic relations remain invariant under linear
transformations. These endomorphisms constitute the quantum supergroup.
 From a rigorous mathematical point of view, the quantum superplane appears
in this approach as a comodule over the corresponding quantum supergroup.
The quantum (super)space has been then envisioned by many as a paradigm for
the general program of quantum deformed physics. The most hopped for
applications includes a possible role in a future quantized theory
of gravity \cite{Maj}.
In this direction many efforts have been accomplished in order to
develop its differential structure \cite{KU92}. But all these constructions
are made in the sense of \cite{WZ90} and not, as far as we know, in the
sense of Woronowicz because of the lack of the Hopf structure of the quantum
super-plane.

More recently, one of us \cite{tahri} has introduced the quantum plane as a quantum group according to the
Faddeev-Reshetikhin-Takhtajan approach \cite{FRT}.
In the present paper we extend the
above investigation to the $1+1$-dimensional quantum super-plane. In this new
approach the quantum super-plane appears as a quantum sub-supergroup
of $SL_q(1/1)$. We give its super-matrix element, its
corresponding R-matrix and its Hopf structure. The latter appears
to be cocommutative. We then realize its dual Hopf superalgebra starting
from postulated initial pairings. We also construct
a right invariant differential
calculus on it and we deduce its corresponding quantum
Lie superalgebra which is as commutation superalgebra,
up to a redefinition of the generators, the
classical $N=1$ supersymmetric algebra, but its Hopf structure
is a non-cocommutative $q$-deformed one. This quantum Lie superalgebra is
isomorphic to the one we obtained by the above duality.

The paper is organized as follows. In section 2, we recall the Manin
approach. In section 3, the $1+1$-dimensional quantum superplane $K_q^{1/1}$
is made in the FRT approach, we also discuss some important properties
of the quantum supermatrix considered to be a group element of $K_q^{1/1}$.
In section 4, we find the dual Hopf superalgebra. In section 5,
we construct a right invariant differential calculus.
In section 6, we derive the quantum Lie superalgebra, for which we give
the Hopf structure in section 7. Finally, we give some concluding remarks
in section 8.

\section{Review of the Manin approach}
The simplest example illustrating this approach is the one parameter quantum
deformation $GL_q(1/1)$ of the supergroup $GL(1/1)$ of $2\times 2$
supermatrices with two bosonic and two fermionic matrix elements.
The construction of this quantum supergroup is well studied in
\cite{Man89} \cite{SSV90} \cite{CFFS}. It is generated by the
elements of a quantum supermatrix:
\beq\label{supmatr}
M~=~\bl(\bra{lcr}
&a&\be\\
&\ga &d
\era\br)
\eeq
which obey the following supercommutation relations:
\beq\label{glq}
\bra{lcr}
&a\be=q\be a,\hfill &d\be=q\be d\hfill\\
&a\ga=q\ga a,\hfill &d\ga=q\ga d\hfill\\
&\be\ga +\ga\be=0,\hfill &\be^2=\ga^2=0\\
&ad-da=-\lambda\be\ga, &\\
\era
\eeq
where $\lambda=q-q^{-1}$ and the generators $a$ and $d$ are even
with parity $\p a=\p d=0$
and $\be$, $\ga$ are odd with parity $\p\be=\p\ga=1$(called Grassmanian).
The supermatrix $M$ defines the basic representation of $GL_q(1/1)$,
i.e., $M$ is considered as a group element. Thus, in this sense, the quantum
supergroup $GL_q(1/1)$ appears as a deformation of the algebra of
polynomials function on $GL(1/1)$, the later is recovered when one
takes $q\to 1$.

The relations (\ref{glq}) may be succinctly expressed
in terms of the graded RTT equation as
\beq\label{rtt}
RM_1M_2=M_2M_1R,
\eeq
where
\beq\label{tensor}
\bra{ll}
&(M_1)^{ij}_{kl}=(M\ot 1)^{ij}_{kl}=(-1)^{(\p k(\p j+ \p l))}M^i_k\de^j_l\\
&(M_2)^{ij}_{kl}=(1\ot M)^{ij}_{kl}=(-1)^{(\p i(\p k +\p l))}M^j_l\de^i_k
\era
\eeq
and the matrix $R$ is given by:
\beq\label{rmatr}
R~=~
\pmatrix{&q&0&0&0\cr
&0&1&0&0\cr
&0&\lambda&1&0\cr
&0&0&0&q^{-1}\cr}
\eeq
which is a solution of the graded quantum
Yang-Baxter equation:
\beq
R_{12}R_{13}R_{23}=R_{23}R_{13}R_{12}
\eeq
arising from the associativity requirement of $GL_q(1/1)$.

In this formulation one can define the quantum superdeterminant:
\beq\bra{lr}
{\cal D}=Sdet(M)&=ad^{-1}-\be d^{-1}\ga d^{-1}\\
&=d^{-1}a-d^{-1}\be\ga d^{-1}\\
\era
\eeq
provided $d^{-1}$ exists, and it can be checked that it is central, i.e.,
it commutes with all elements of $M$. Therefore, by imposing
the relation ${\cal D}=1$, we may define a quantum deformation
$SL_q(1/1)$ by analogy with the classical restriction to the special
linear supergroup. The inverse supermatrix can be now defined.
It is given by
\beq \label{inverse}
M^{-1}=
\pmatrix{
&a^{-1}+a^{-1}\be d^{-1}\ga a^{-1} & -a^{-1}\be d^{-1}\cr
&{-d^{-1}\ga a^{-1}}&d^{-1}+d^{-1}\ga a^{-1}\be d^{-1}\cr
}.
\eeq
So the Hopf structure of $GL_q(1/1)$ is given by the following:
\beq\label{hopfalg}
\De(M)=M\dot{\ot }M=
\pmatrix{
& a\ot a+\be\ot\ga &a\ot\be+\be\ot d\cr
&\ga\ot a +d\ot\ga&\ga\ot \be+d\ot d\cr
}
\eeq
\beq\label{counit}
\vep(M)=
\pmatrix{
&1&0\cr
&0&1\cr
}
\eeq
\beq
S(M)=M^{-1}.
\eeq

Now let us recall some statements related to the supermatrix $M$
considered as group element (or a point) of $GL_q(1/1)$.
\nt If one takes two copies $M$ and $M'$ such that their elements pairwise supercommute, then the products $MM'$ and $M'M$ are group elements
of $GL_{q}(1/1)$. Furthermore, the inverse $M^{-1}$ is
actually a group element of $GL_{q^{-1}}(1/1)$ rather than $GL_q(1/1)$,
and, in general, the n-th power of $M$,
$n\in {\ZZ}$, is a group element of $GL_{q^n}(1/1)$.

Manin has defined in correspondence with the quantum
supergroup what is called the quantum superplane or `quantum superspace'
and the dual of it. This quantum superplane or, rather, the polynomial
function ring on it, denoted by $K_q^{1/1}$, is generated
by two coordinates $x$ bosonic with parity $\p x=0$ and $\te$
fermionic with parity $\p\te=1$ such that:
\beq\label{supplane}\bra{llrr}
&x\te~=~q\te x&\qquad\qquad(q\neq 0,1)\\
&\te^2~=~0.&\\
\era
\eeq
The dual, denoted by ${}^*K_q^{1/1}$, is generated by coordinates
$\xi$ with parity $1$ and $y$ with parity $0$ and the following
supercommutation relations:
\beq\label{dqsp}\bra{llrr}
&\xi y~=~q^{-1}y \xi&\qquad\qquad(q\neq 0,1)\\
&\xi^2~=~0.&\\
\era
\eeq
The quantum supergroup $GL_q(1/1)$ appears as
endomorphisms of the quantum superplanes
$K_q^{1/1}$ and ${}^*K_q^{1/1}$ under which the quadratic relations
(\ref{supplane}) and (\ref{dqsp}) remain invariant. So Manin proved
the equivalence between the defining quantum supergroup supercommutation
relations (\ref{glq}) and the quadratic relations (\ref{supplane}) and
(\ref{dqsp}).
In a more rigorous mathematical point of view, the quantum superplanes
$K_q^{1/1}$ and ${}^*K_q^{1/1}$ are viewed as comodules over $GL_q(1/1)$
via the following (left) coactions
\beq\label{lcoac}
\de(X)=M\dot{\ot }X=
\pmatrix{
& a\ot x+\be\ot\te \cr
&\ga\ot x +d\ot\te \cr
}
\eeq
\beq\label{lcoad}
\de(\Te)=M\dot{\ot }\Te=
\pmatrix{
& a\ot \xi+\be\ot y \cr
&\ga\ot \xi +d\ot y \cr
}
\eeq
where $X$ and $\Te$ are column vectors of components $x, \te$ and
$\xi, y$, respectively.

\section{Quantum superplane in new approach.}
Let us consider a particular copy of $GL_q(1/1)$ where
$a=d=x$ and $\be=\ga=\te$
that is:
\beq\label{matra}
A~=~\pmatrix{
&x &\te\cr&\te&x\cr},
\eeq
in this case the commutation relations (\ref{glq}) reduce
to the quadratic relations (\ref{supplane}) defining the quantum superplane
$K^{1/1}_q$. Note that the superdeterminant of $A$ is equal to $1$,
so it may be viewed also as a particular copy of $SL_q(1/1)$.
Its quantum inverse supermatrix is given by
\beq\label{inva}
A^{-1}=
\pmatrix{
&x^{-1}&-x^{-1}\te x^{-1}\cr
&-x^{-1}\te x^{-1}& x^{-1}\cr}.
\eeq
This leads to our following main result.\\\\

\nt{{\bf Thoerem.}\\{\it
The quantum Manin superplane $K_q^{1/1}$ generated by the coordinates
$x$ with parity $0$ and $\te$ with parity $1$ and extended by
the inverse $x^{-1}$ satisfying the quadratic relations
$x\te=q\te x$ and $\te^2=0$ is a quantum supergroup generated by
elements of the supermatrix of type $A$ modulo the graded $RTT$
equation (\ref{rtt}) where the matrix $R$ is given by (\ref{rmatr}).
Its Hopf structure is given explicitly by:
\beq\label{hopfsup}\bra{lllll}
&\De(x)=x\ot x+\te \ot\te,\\
&\De(\te)=\te\ot x +x\ot\te,\\
&\vep(x)=1,\\
&\vep(\te)=0,\\
&S(x)=x^{-1}\\
&S(\te)=-x^{-1}\te x^{-1}.
\era\eeq
}
We see that it is a cocommutative Hopf superalgebra.

There are curious properties applicable to this quantum supermatrix
 considered as a group element of $K_q^{1/1}$. They assert that:\\
a) If $A$ and $A'$ are two copies of $K_q^{1/1}$ such that their
entries pairwise supercommute, then the products
$AA'$ and $A'A$ are group elements of $K_q^{1/1}$.\\
b) the n-th power of $A$ is a group element of $K_{q^{n}}^{1/1}$ given by:
\beq{\label{pmata}}
A^n=\pmatrix{
&x^n&[n]_q\te x^{n-1}\cr
&[n]_q\te x^{n-1}&x^n\cr},
\eeq
where $[n]_q=(1-q^n)/(1-q)$. Note that if we take $n=-1$ we find the
inverse (\ref{inva}) (since $[-1]_q=-q^{-1}$). Furthermore, if there
exists an integer $n$ such that $q^n=1$, then the n-th power of $A$
(\ref{pmata}) reduces to
\beq{\label{rpmat}}
A^n=\pmatrix{
&x^n&0\cr
&0&x^n\cr},
\eeq
which gives a nilpotent supermatrix when $x^n=0$ or unit supermatrix
when $x^n=1$.

So in this sense, the quantum superplane $K_q^{1/1}$ can be considered as
a quantum subsupergroup of $SL_q(1/1)$.

\section{The dual of $K_q^{1/1}$}
We will apply here the method introduced by Sudbery in \cite{Sud}, where
he obtained ${{\cU}_q}(sl(2))\ot {\cU}(u(2))$ as the algebra of tangent vectors
at the identity, and generalized to more complicated algebras by Dobrev
\cite{Dob1}. The application of this method to the multiparameter
supergroup $GL_{u{\bf q}}(m/n)$ is given in \cite{dota}. It consists of
using the duality between (super)bialgebras or Hopf (super)algebras to
obtain the unknown dual of a known (super)algebra.

Two bialgebras $\cU$ and $\cA$ are said to be in duality \cite{Abe}
if there exists a doubly nondegenerate bilinear form
\beq{\label{dubi}}
\lag~,~\rad :~{\cU\times\cA\longrightarrow\CC} \quad  (u,a)\lpt\lag~u,a~\rad
\quad {u\in \cU},~a\in\cA
\eeq
such that, for $u,~v\in {\cU}$, $a,~b\in\cA$
\beq\label{prdu}
\bra{ll}
&\lag~u,ab~\rad =\lag~\de_{\cU}(u),a\ot b~\rad ,\\
&\lag~uv,a~\rad =\lag~u\ot v,\de_{\cA}(a)~\rad ,\\
&\lag~u,1_{\cA}~\rad =\vep_{\cU}(u),\quad
\lag~1_{\cU},a~\rad =\vep_{\cA}(a).\\
\era
\eeq
For dual Hopf algebras one adds
\beq{\label{dhal}}
\lag~\ga_{\cU}(u),a~\rad~=~\lag~u,\ga_{\cA}(a)~\rad.
\eeq
To extend this to superalgebras or Hopf superalgebras, we take into account
that the tensor product is also graded, and, if (using Sweedler's notation)
$\de_{\cU}=\sum u_{(1)}\ot u_{(2)},~\de_{\cA}=\sum a_{(1)}\ot a_{(2)}$, then
\beq\label{grdu}
\bra{ll}
&\lag~u,ab~\rad =\sum (-1)^{\widehat{u_{(2)}}\widehat{a}}
\lag~u_{(1)},a~\rad\lag~u_{(2)},b~\rad ,\\
&\lag~uv,a~\rad =\sum (-1)^{\widehat{v}\widehat{a_{(1)}}}
\lag~u,a_{(1)}~\rad\lag~v,a_{(2)}~\rad ~.\\
\era
\eeq
It is enough to define the pairing (\ref{dubi}) between the generating
elements of the two algebras. The pairing may then be extended to all
elements of $\cU$ and $\cA$ by using relations (\ref{prdu}).

To find the unknown superalgebra, it is enough to give the pairing
between the generating elements of the unknown superalgebra with an arbitrary
element of the PBW basis of the known superalgebra. So, we first need
to fix a PBW basis of $K_q^{1/1}$. This basis consists of monomials
\beq{\label{pbwb}}
h=x^m\te^n
\eeq
where $m\in{\ZZ}$ and $n\in\{0,1\}$. Let us then denote the dual
superalgebra by ${\cU}_q(k(1/1))$ and its generating elements by
$\chi$ and $\phi$. Following \cite{Dob1} we shall postulate the
pairing $\lag Z,h\rad$ $Z=\chi,~\phi$, $h$ from (\ref{pbwb}), as
we use the classical tangent vector at the identity :
\beq\label{tvai}
\bra{ll}
&\lag~\chi,h~\rad =m\de_{n,0},\\
&\lag~\phi,h~\rad =\de_{n,1}.\\
\era
\eeq
We also note that, from the duality properties, it follows
\beq{\label{duco}}
\lag~1_{\cU},h~\rad =\vep(h)=\de_{n,0}.
\eeq
To obtain the commutation relations between the generators $\chi$ and
$\phi$, we first need to evaluate the action of their bilinear
product on the elements of $K_q^{1/1}$. Using the defining relations
we obtain
\beq\label{bpro}
\bra{ll}
&\lag~\chi\phi,h~\rad =(m+1)\de_{n,0},\\
&\lag~\phi\chi,h~\rad =(m+1)\de_{n,1},\\
&\lag~(\phi)^2,h~\rad =-{\frac{1-q^{-2m}}{1-q^{-2}}}\de_{n,0}.\\
\era
\eeq
Thus we have the commutation relations :
\beq\label{cxph}
\bra{ll}
&\chi\phi~-~\phi\chi ~=~0,\\
&(\phi)^2~=~-\frac{1-q^{-2\chi}}{1-q^{-2}}.\\
\era
\eeq
The Hopf structure of this superalgebra may be deduced by using the
duality. We start with the coproduct, i.e., we shall use the relations
\beq{\label{dcop}}
\lag~Z,h~\rad =\lag~\de_{\cU}(Z),h_1\ot h_2~\rad
\eeq
for every splitting $h=h_1h_2$. Thus we get
\beq{\label{coda}}
\bra{ll}
&\de_{\cU}(\chi)~=~\chi\ot 1~+~1\ot\chi,\\
&\de_{\cU}(\phi)~=~\phi\ot q^{-\chi}~+~1\ot\phi~.\\
\era
\eeq
the counit relations in ${\cU}_q(k(1/1))$ are given by
\beq{\label{cocu}}
\vep_{\cU}(Z)~=~0,\quad Z=\chi,~\phi
\eeq
which follows easily from
\beq{\label{duep}}
\lag~Z,1_{\cA}~\rad ~=~0~=~\vep_{\cU}(Z).
\eeq
Finally, the antipode map of ${\cU}_q(k(1/1))$ follows from
\beq{\label{duan}}
\lag~\ga_{\cU}(Z),h~\rad ~=~\lag~Z,\ga_{\cA}(h)~\rad.
\eeq
It is given by
\beq{\label{anda}}
\bra{ll}
&\ga_{\cU}(\chi)~=~-\chi,\\
&\ga_{\cU}(\phi)~=~-q^{\chi}\phi~.\\
\era
\eeq

Now let us make the following redefinition:
\beq{\label{redp}}
\phi'~=~iq^{(\chi -1)/2}\phi,\quad i^2=-1
\eeq
then we get the following quantum deformed enveloping superalgebra
(dropping the primes)
\beq\label{ncxp}
\bra{ll}
&\chi\phi~-~\phi\chi ~=~0,\\
&(\phi)^2~=~[\chi]\\
\era
\eeq
where $[x]=(q^{x}-q^{-x})/(q-q^{-1})$. Its Hopf structure is now given by:
\beq{\label{nhst}}
\bra{ll}
&\de_{\cU}(\chi)~=~\chi\ot 1~+~1\ot\chi,\\
&\de_{\cU}(\phi)~=~\phi\ot q^{-\chi/2}~+~q^{\chi/2}\ot\phi~.\\
&\vep_{\cU}(\chi)~=~\vep_{\cU}(\phi)~=~0,\\
&\ga_{\cU}(\chi)~=~-\chi,\quad \ga_{\cU}(\phi)~=~-\phi~.\\

\era
\eeq

\section{Differential calculus on  $K_q^{1/1}$}
The differential calculus on quantum group was initiated by
Woronowicz \cite{Wor87} \cite{Wor89}.
Here we follow the same techniques used in \cite{SWZ91}
to construct the right invariant calculus on the two-parameter
deformation of $GL(2)$ and extended in \cite{SVZ90} and
\cite{BT92} to the one and two-parameter deformation of $GL(1/1)$,
respectively.

We introduce at first the exterior differential $d$ which satisfies the
nilpotency and the graded Leibniz rule,
\beq\label{leibr}
\bra{ll}
&d^2=0,\\
&d(fg)=(df)g~+~(-1)^{\p f}f(dg),\\
\era
\eeq
where $f$, $g$ are functions of the variables $x$, $\te$ and ${\p f}$
is the parity of $f$. In analogy with the classical differential geometry,
one can construct from $A$ and the one-forms $dA$, via the matrix valued one-forms $\Om$ , the right-invariant Cartan-Maurer forms as
\beq\label{cmf}
\Om=(dA)A^{-1}=
\Bl(
\bra{llrr}
&\om&v\\
&v&\om
\era
\Br).
\eeq
Writing this as $dA=\Om A$, one can express the basic one-forms
$dx$ and $d\te$ in terms of Cartan-Maurer forms as
\beq\label{bof}
\bra{llrr}
&dx=\om x +v\te\\
&d\te=v x+\om\te.\\
\era
\eeq
The exterior differential $d$ can then be written as
\beq\label{exod}
\bra{llrr}
d~&=~dx\pa_x +d\te\pa_{\te}\\
&=~\om D_x+vD_{\te}.\\
\era
\eeq

Using the nilpotency of the differential
$d$ and (\ref{bof}) it follows that:
\beq\label{dcmf}\bra{l}
d\om=\om^2-v^2\\
dv=\om v -v\om \\
\era
\eeq
which we call the Cartan-Maurer equations.

Now we want to calculate the $q$-deformed supercommutation relations between
the basic variables $x$, $\te$ and the Cartan-Maurer forms. Using a similar
restriction as in \cite{SVZ90} due to the right-invariance of the
Cartan-Maurer forms, we suggest
\beq\label{crvof}\bra{llllr}
&xv=qvx,&\te v=qv\te,&\\
&x\om=F\om x,&\te\om=-F\om\te,
\era
\eeq
where $F$ is an arbitrary complex number. Then the action of $d$
on the quadratic relations (\ref{supplane}) do not lead to
any inconsistency, while the consistence of $d$ with (\ref{crvof})
requires the following:
\beq\label{module1}\bra{llllr}
&\om v= v\om& & \\
&\om^2= kv^2,& &k={\frac{q^2-F}{F(F+1)}}.
\era
\eeq
Using this, the Cartan-Maurer equations (\ref{dcmf}) reduce to
\beq\label{rdcmf}\bra{l}
d\om=(k-1)v^2\\
dv=0\\
\era
\eeq

\section{Quantum Lie superalgebra }

Now let us return to the expression of $d$ in (\ref{exod}), the
elements $D_x$ and $D_{\te}$ generate the quantum Lie superalgebra
associated to the quantum superplane $K_q^{1/1}$. To calculate the
supercommutation relations between these generators we use the nilpotency
of $d$,

\beq\label{niof}\bra{ll}
0=d^2(f) &=~d(\om D_xf +vD_{\te}f)\cr
&=~(d\om)D_xf+(dv)D_{\te}f-\om(\om D_x+vD_{\te})D_xf+\cr
&+~v(\om D_x+vD_{\te})D_{\te}f,\cr
\era
\eeq
where $f$ is an arbitrary function of the variables $x$, $\te$.
Then, using (\ref{module1}) and (\ref{rdcmf}), it follows that
\beq\label{algebre1}
\bra{ll}
&D_xD_{\te} -D_{\te} D_x=0\\
&D_{\te}^2=kD_x^2+(1-k)D_x.
\era
\eeq
We will denote this superalgebra by $k_{q,F}(1/1)$. Note that
if we make the following change:
\beq\label{ndx}
T_x=kD_x^2 +(1-k)D_x,
\eeq
then we get
\beq\label{algebre2}
\bra{ll}
&T_xD_{\te} -D_{\te} T_x=0,\\
&D_{\te}^2=T_x,
\era
\eeq
which are the commutation relation of the `classical' $N=1$
supersymmetric algebra. So the
quantum deformation does not affect the Lie superalgebra structure.

Now it is possible to calculate the supercommutation relations between
the generators of $k_{q,F}(1/1)$ and the basic coordinates.
To this end, we use the graded Leibniz rule by comparing the coefficients
of the Cartan-Maurer forms when acting by $d$ on $xf$ or $\te f$.
Then we obtain
\beq\label{module2}\bra{llrr}
&D_xx=x+FxD_x,&D_x\te =\te+F\te D_x\\
&D_{\te} x=\te +qxD_{\te},&D_{\te}\te=x-q\te D_{\te}.
\era
\eeq
On the other hand, if we multiply the second relation in (\ref{algebre1})
from the right by $x$ or $\te$ and then pull it to the left using the
commutation relations (\ref{module2}), then we get
\beq\label{ccod}
D_{\te}^2-\frac{kF^2}{q^2}D_x^2-\frac{F(1+k)}{q^2}D_x=0
\eeq
from which it follows that $F$ must be equal to
$q^2$ or $q$. We will see in the next section that these
two solutions are equivalent.

\section{Hopf structure of the superalgebra $k_{q,F}(1/1)$}
Now we shall use the graded Leibniz rule for arbitrary
functions $f$ and $g$ of coordinates $x$, $\te$ to derive the Hopf structure
of $k_{q,F}(1/1)$. An arbitrary function is
in fact a linear combination of monomials in the basis (\ref{pbwb}).

We first need to calculate the commutation relations
between the Cartan-Maurer one-forms and functions. From (\ref{crvof})
we have
\beq\label{ccmf}
\bra{ll}
&f\om=(-1)^{\p f}wF^Nf\\
&fv=vq^Nf.
\era
\eeq
where $N$ is an operator number acting on monomial as
\beq\label{opnu}
N(x^m\te^n)=(m+n)x^m\te^n,
\eeq
and may be related to the operator $D_x$. To see this we calculate
the action of $D_x$ on functions which may be achieved by calculating its
commutation relation with $h$. Indeed we have
\beq\label{crwh}
D_{x}h=\frac{1-F^{m+n}}{1-F}h~+~F^nhD_x
\eeq
from which we get
\beq\label{rdxn}
D_x=\frac{1-F^N}{1-F}
\eeq

Therefore, using the graded Leibniz rule applied to the product of functions
$f$ and $g$, and the expression of $d$ in terms of $D_x$ and $D_{\te}$
(\ref{exod}) with the help of (\ref{ccmf}), this leads to the following:
\beq\label{dofg}\bra{ll}
&D_x(fg)=D_x(f)g+F^NfD_x(g),\\
&D_{\te}(fg)=D_{\te}(f)g+(-1)^{\p f}q^NfD_{\te}(g),\\
\era
\eeq
from which we get the coproduct:
\beq\label{copk}
\bra{ll}
&\de(D_x)=D_x\ot 1+F^N\ot D_x\\
&\de(D_{\te})=D_{\te}\ot 1+q^N\ot D_{\te}
\era
\eeq
where use was made of the graded tensor product: $(a\ot b)(f\ot g)=
(-1)^{{\p b}{\p f}}a(f)\ot b(g)$, $a, b\in  k_{q,F}(1/1)$
and $f, g\in K_q^{1/1}$. The counit and antipode relations in
$k_{q,F}(1/1)$ may be calculated by using the basic axioms
of Hopf superalgebras \cite{Abe} \cite{Man89}:
\beq{\label{axha}}
\bra{ll}
(\vep_k\ot id)\de_k(u)~=~u,\\
m(id\ot\ga_k)\de_k(u)~=~\vep_k(u),\\
\era
\eeq
where $u\in k_{q,F}(1/1)$ and $m$ is the multiplication.
So we obtain
\beq{\label{canr}}
\bra{ll}
&\vep_k(D_x)~=~\vep_k(D_{\te})~=~0,\\
&\ga_k(D_x)~=~-F^{-N}D_x,\quad \ga_k(D_{\te})~=~-q^{-N}D_{\te}.\\
\era
\eeq
Thus, we have a $q$-deformed non-cocommutative Hopf
structure depending on the parameter $q$. Furthermore,
we note that the two forms of (\ref{copk}), and (\ref{canr})
according to the two solutions of $F=q^2$ or $q$ are consistent
with the redefinition (\ref{ndx}). Therefore, we may view the two solutions
of $F$ as corresponding to changes of basis for the quantum enveloping
superalgebra and are equivalent as such. So, it is convenient to choose
the solution $F=q^2$ which is the simplest one. We can then transform
this superalgebra to the form obtained in Section 4 (c.f., (\ref{ncxp})
(\ref{nhst})) by making the following changes:
\beq\label{comp}
\bra{ll}
&D_x~\equiv~{\frac{1-q^{2\chi}}{1-q^2}}~,\\
&\\
&D_{\te}~\equiv~q^{(\chi-1)/2}\phi\\
\era
\eeq
which are consistent with the commutation relations and the Hopf structure.

\section{Concluding remarks }
We have shown that the 1+1-dimensional quantum superplane, which is
well known in the literature as comodule over its corresponding
quantum supergroup, may be incorporated in the Faddeev-Reshetikhin-Takhtajan
approach to quantum (super)groups. We have given its supermatrix element,
its corresponding R-matrix and its Hopf structure which
is a cocommutative one. We then derived its dual Hopf superalgebra starting
from a postulated initial pairing. We also constructed a right
differential calculus on it and we deduced the corresponding quantum
Lie superalgebra. As a commutation superalgebra, the latter coincides with
the classical N=1 supersymmetric algebra, but its Hopf structure is a
non-cocommutative $q$-deformed one. We also prove that there is an
isomorphism between this quantum Lie superalgebra and the one we obtained
using postulated duality.
To finish let us stress that the concept of $q$-deformation of Lie
(super)group and (super)space is physically consistent when it is
performed in the context of Hopf (super)algebra. So we expect that
this new approach to the quantum superplane and the differential
calculus introduced on it here will be a starting point to construct
more consistent $q$-deformed physical models.

{\vskip 8mm
{\bf Acknowledgments.}~~~
One of the authors MEF is grateful for hospitality at the Abdus Salam ICTP
, where part of this work was achieved, to Prof. R. Floreanini, and to Prof. S. Randjbar-Daemi for reading this paper.  }

\end{document}